# Educational Objectives Of Different Laboratory Types:
# A Comparative Study

Yasser .H. Elawady

Faculty of computers and Information systems
Taif University,
Taif, KSA.
y.alawadi@tu.edu.sa

A.S. Tolba

Faculty of Computer Studies
Arab Open University,
HQ, Kuwait.
a.tolba@arabou.edu.kw

*Abstract*— Lab courses play a critical role in scientific education. Modern technology is changing the nature of the laboratories, and there is a great comparision between hands-on, simulated and remote laboratories. The remote lab technology has brought a significant improvement in communication within the Academic community and has improved students' learning experiences. There are different educational objectives as criteria for judging the laboratories: Hands-on labs increase the Ability to design and investigate (design skills), while remote labs focus on conceptual understanding. Remote laboratories offer all the advantages of the new technology, but are often a poor replacement for real laboratory work. Remote laboratories are similar to simulation techniques in that they require minimal space and time, because the experiments can be rapidly configured and run over the Internet [Web]. But unlike simulations, they provide real data. This paper presents a comparative analysis for the educational objectives of the three laboratory techniques; hands-on, simulated, and remote laboratories. In addition, it proposes enhancements for the remote lab activities leading to improving its performance.

## I. INTRODUCTION

Different educational objectives are used as criteria for judging the laboratories: Hands-on advocates emphasize design skills, while remote lab advocates focus on conceptual understanding. Nersessian [1991] goes so far as to claim that "hands-on experience is at the heart of science learning" and Clough [2002] declares that laboratory experiences "make science come alive." Lab courses have a strong impact on students' learning outcomes, according to Magin et al. [1986].

This domain of study ranges across many disciplines, and is challenging to survey. In order to find the existing literature, we focused on three electronic databases: ACM, IEEE, and Science Direct. As a result, 60 articles were selected for a full-text review and coding ( 20 publications for each; hands- on labs, simulated labs, and remote labs). *These articles are listed in the Appendix.*

Most of the literature focuses on engineering laboratories as the engineering discipline contains the biggest portion of laboratory studies. Engineering professors may also see the labs as connected to future employment [Faucher 1985]. In other words, engineering is an applied science. Alternatively, the impetus for the creation of a remote laboratory may come from an engineer's desire to build something. This paper presents a comparative analysis for the educational objectives of the three laboratory techniques; hands-on, simulated, and remote laboratories. In addition, it proposes enhancements for the remote lab activities leading to improving its performance.

The rest of this paper is organized as follows: Section II introduces Comparison of Different Laboratory Types. Section III introduces analysis and discussion of the educational Objectives. Section IV presents our conclusion. Finally, section V concludes our Recommendations for enhancing the performance of remote laboratories.

## II. COMPARISON OF DIFFERENT LABORATORY TYPES

The three types of labs are sometimes compared to each other, while in other cases the labs are merged. The integrated teaching and learning (ITL) program at the University of Colorado at Boulder provided an example of how to combine hands-on practice with simulation experience and remote experimentation [Schwartz and Dunkin 2000]. A handful of articles evaluated remote laboratories in comparison to hands-on laboratories [Sicker et al. 2005] or simulated laboratories in comparison to hands-on laboratories [Engum et al. 2003]. Engum et al. [2003] showed that hands-on labs were more effective than simulated.

A summarized description of the three types of labs is described below.

- **Hands-On Labs**: Hands-on labs involve a physically real investigation process. Two characteristics distinguish hands-on from the other two labs: (1) All the equipment required to perform the laboratory is physically set up; and (2) the students who perform the laboratory are physically present in the lab. On the other hand, hands-on experiments are seen as too costly. Hands-on labs put a high demand on space, instructor time, and experimental infrastructure, all of which are subject to rising costs [Farrington et al. 1994] Also, due to the limitation of space and resources, hands-on labs are unable to meet some of the special needs of disabled students [Colwell et al. 2002] and distant users [Watt et al. 2002]. Additionally, students' assessments suggest that students are not satisfied with current hands-on labs [Dobson et al. 1995].

- **Simulated Labs**: Simulated labs are the imitations of real experiments. All the infrastructure required for laboratories is not real, but simulated on computers. Some note that the cost of simulation is not necessarily lower than that of real labs [Canizares and Faur 1997]. Realistic simulations take a large amount of time and expense to develop and still may fail to faithfully model reality [Papathanassiou et al. 1999]. The theory of situated learning (e.g., McLellan [1995]) would suggest that what students learn from simulations is primarily how to run simulations.

- **Remote Labs:** Remote labs are characterized by mediated reality. Similar to hands-on labs, they require space and devices. What makes them different from real labs is the distance between the experiment and the experimenter. In real labs, the equipment might be mediated through computer control, but collocated. By contrast, in remote labs experimenters obtain data by controlling geographically detached equipment.

In other words, Reality in remote labs is mediated by distance. Remote labs are becoming more popular [Yoo and Hovis 2004]. They have the potential to provide affordable real experimental data through sharing experimental devices with a pool of schools [Sonnenwald et al. 2003]. Also, a remote lab can extend the capability of a conventional laboratory. Along one dimension, its flexibility increases the number of times and places a student can perform experiments [Canfora et al. 2004].

Along another, its availability is extended to more students [Cooper et al. 2002b]. Additionally, comparative studies show that students are motivated and willing to work in remote labs [Cooper et al. 2002b]. Some students even think remote labs are more effective than working with simulators [Scanlon et al. 2004].

## III. ANALYSIS AND DISCUSSION OF THE EDUCATIONAL OBJECTIVES

In order to study this hypothesis, first, the articles (in the appendix) are coded based on educational objectives. A four-dimensional goal model is developed for laboratory education (see Table I). This model is built starting with the educational goals proposed by the Accreditation Board for Engineering and Technology (ABET) [2005].

Table I. Educational Goals for Laboratory Learning

| Lab Goals | Description | Goals from ABET |
|---|---|---|
| Conceptual understanding | Extent to which laboratory activities help students understand and solve problems related to key concepts taught in the classroom. | Illustrate concepts and principles. |
| Design skills | Extent to which laboratory activities increase student's ability to solve open-ended problems through the design and construction of new artifacts or processes. | Ability to design and investigate. Understand the nature of science (Scientific mind). |
| Social skills | Extent to which students learn how to productively perform engineering-related activities in groups. | Social skills and other productive team behaviors (communication, team interaction and problem solving, leadership). |
| Professional skills | Extent to which students become familiar with the technical skills they will be expected to have when practicing in the profession. | Technical/procedural skills. Introduce students to the world of scientists and engineers in practice. Application of knowledge to practice. |

Figures 1, 2 and 3 shows the educational goals for each lab summarized from the articles reviewed.

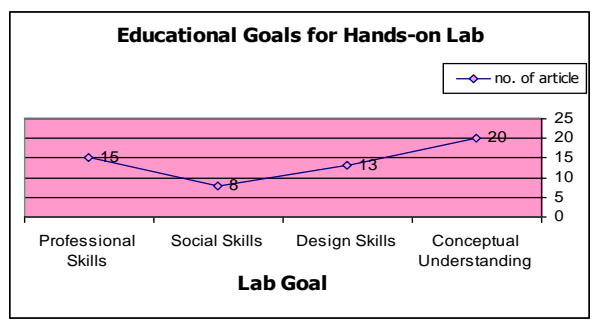

Figure 1. Educational goals for hands-on lab

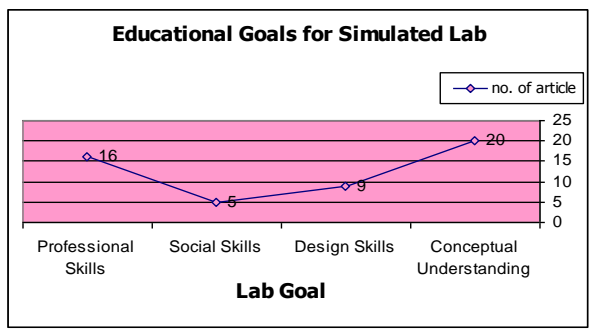

Figure 2. Educational goals for simulated lab

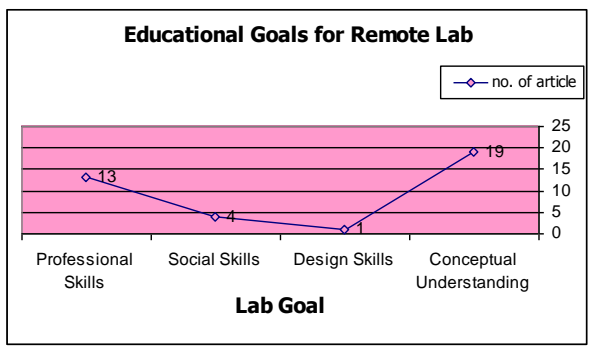

Figure 3. Educational goals for Remote lab

Figure 1. shows that, for hands-on labs, all four educational goals are well-addressed by most of the articles.In particular, the literature on hands-on labs placed a strong emphasis on conceptual understanding and design skills. Professional skills were also recognized as an important mission for hands-on labs. We found that more than half of the hands-on laboratory articles recognized the importance of design skills and agreed that design skill is an important goal for hands-on labs. The dimension of social skills is least represented in the articles. They are not discussed as often as other educational goals.

Figure 2. shows that, Articles on simulated laboratories are skew even more towards conceptual understanding and professional skills. All the articles discuss conceptual understanding; less than half address design skills.

Figure 3. shows that, Articles on Remote laboratory are dramatically different. They focus on conceptual understanding and professional skills. Only one work discusses design skills. This is an interesting result. It suggests that the proponents of remote laboratories may think of their success only in reference to conceptual and professional learning. It may be that they do not think remote laboratories are appropriate for teaching design skills.

*Our research will contribute in enhancing both social and design skills by augmenting the remotely accessed lab with the second life.*

**Second Life** (**SL**) is a virtual world developed by Linden Lab launched on June 23, 2003 and is accessible via the Internet. A free client program called the Second Life Viewer enables its users, called Residents, to interact with each other through avatars. Residents can explore, meet other residents, socialize, participate in individual and group activities, and create and trade virtual property and services with one another, or travel throughout the world, which residents refer to as the grid.[26]

The results from this sample of articles suggest the following possible explanation for the debate over laboratories. Adherents of hands-on laboratories find other laboratories to be lacking. They do not believe that alternative labs can be used in teaching design skills. By contrast, adherents of remote laboratories think the hands-on laboratory researchers are ignoring evidence which shows that remote laboratories are effective in teaching concepts. Remote laboratory adherents are evaluating their own efforts with respect to teaching concepts, not design skills.

On the other hand, researchers are confounding many different factors, and perhaps over-attributing learning success to the technologies used. There is much in the literature to suggest that both students' preferences and learning outcomes are the result of many intertwined factors. Thus, it is sensible to suggest that researchers more carefully isolate and study the different factors which might interact with laboratory technology in determining educational effectiveness. However, such work is difficult. It is hard to perform large-scale educational tests and hold factors such as instructor ability constant. It is also difficult to compare studies which focus on different scientific domains. Thus, it is especially important that effort should be focused on areas that look the most promising.

- **First**, research may look at hybrids of laboratories that are designed to accomplish a portfolio of educational objectives. There is a fair amount of evidence that simulated and remote labs are effective in teaching concepts.

- **Second,** the effectiveness of laboratories may be affected by how much students believe in them. Therefore, an understanding of presence, interaction, and belief may lead to better interfaces.

- **Third,** *research* might pay more attention to collaboration and sense making. The technology may change the way we can and should coordinate our work.

**The main advantages and disadvantages of each type of laboratory according to some features are summarized in Table II.**

Table II.   Comparative list of advantages and disadvantages of Real, Virtual and Remote laboratories.

| Feature | Hands-on Labs | Simulated Labs | Remote Labs |
|---|---|---|---|
| **Access Mode** | Physical access to lab. | Virtual access to experiments using simulation programs. | Using the internet and SW to access the lab remotely. |
| | Adv. | Adv. | Adv. |
| | • Realistic data.<br>• Interaction with real equipment.<br>• Open ended experiments are possible. | • Good for concept validation.<br>• No time and physical restrictions. | • No time and space restrictions.<br>• Realistic data.<br>• Feeling of reality. |
| | Disadv. | Disadv. | Disadv. |
| | • Limited to available physical environment.<br>• Inflexible lab room ( needs schedules). | • Idealized data.<br>• No interaction with real equipment. | • Only virtual presence in the lab. |
| **Infrastructure** | HW components and computers if required. | Simulation SW programs. | Hardware components, computers and communication media. |
| | Adv. | Adv. | Adv. |
| | • Offer students the sense of the reality.<br>• Help students to connect the experiment under staff supervision. | • Good for conceptual understanding.<br>• Secure if safety precautions are taken into account. | • Offer students to make the experiment more times.<br>• Useful if more real results are required. |
| | Disadv. | Disadv. | Disadv. |
| | • Finite lifetime of the HW components.<br>• Needs maintenance of the HW components.<br>• Vulnerable to damage ( misuse, theft,……). | • Need SW update. | • Finite lifetime of the HW components.<br>• Needs maintenance of the HW components.<br>• Communication problems. |
| **Pedagogical** | Adv. | Adv. | Adv. |
| | • Nothing like real experiment.<br>• Interaction with supervisor.<br>• Offer students to collaborate.<br>• Offer students to learn by trial and error. | • Have more Pedagogical adv. than other labs.<br>• Provide safe learning environment for experimentation with dangerous equipment.<br>• Flexible and easy to use SW tools.<br>• Enhancement through animation and virtual reality software. | • Feeling with reality in data.<br>• Suitable for distance learning.<br>• Focus on conceptual understanding and professional skills. |
| | Disadv. | Disadv. | Disadv. |
| | • Students may not complete experiments in lab period.<br>• Supervision required. | • Supervision of academic staff not available.<br>• No sense with real equipment of the experiment . | • Need enhancing in both social and design skills. |
| **Economical** | • Expensive (Disadv.) | • Low cost (Adv.) | • Medium cost if reduces the number of used labs. (Adv.) |

## IV.   CONCLUSION

We found that most of the articles discussing the educational objectives of different laboratory types were engineering-related. Additionally, there were advocates and detractors for each different type of laboratory. We asked what might explain the continued unresolved debate. The debate can be partially explained by examining the educational objectives associated with each laboratory type. Hands-on lab adherents emphasize the acquisition of design skills as an important educational goal, while remote laboratory adherents do not evaluate their own technology with respect to this objective.

In conclusion, there is no simple answer to the question, which laboratory is the best for engineering students? All types of laboratories offer certain advantages. We believe that engineering students should be offered through the duration of their programs a balanced mixture of real, virtual and remote labs.

**This paper provides** a starting place for researchers involved in the discussion about the role and value of laboratory work. Perhaps a sense of reality can be achieved by students not only in hands-on experience, but also in virtual environments. It is sure with the proper mix of technologies we can find solutions that meet the economic constraints of laboratories by using simulations and remote labs to reinforce conceptual understanding, while at the same time providing enough open-ended interaction to teach design. Our review suggests that there is room for research that seeks to create such a mix, which might be informed by studies of coordination as well as the interactions that lead students to a sense of immersion.

## V. RECOMMENDATIONS FOR ENHANCING THE PERFORMANCE OF REMOTE LABS

- Improving social skills through constructing distributed remote labs.
- Improving design skills through constructing remote labs for the applications which basically depend on computers in real labs such as FPGA labs and other related labs.
- Development of Augmented Reality Labs. Augmented Reality Laboratories (ARLs) Combined Remote Lab Access with Second Life. Adobe Conferencing system with Webcam for lab visualization through the Web in Second life.

## APPENDIX

### I. TABLES OF EDUCATIONAL OBJECTIVES AND ARTICLES

| | |
|---|---|
| Acc.&Flex.: Accessibility & Flexibility | Phy: Physiology |
| Art.: Article<br>**Subject (sub.)**<br>AE: Aeronautical Engineering | PE: Power Engineering<br>SE: Science and Engineering (physics, biology, EE) |
| B: Biology | TE: Telecommunication |
| CE: Chemical Engineering (chemistry)<br>Clm: Climatology<br>CS: Computer Science<br>CVE: Civil Engineering | **Methodology (meth.)**<br>Empirical—variance-based methods<br>Qualitative—conceptualization and evaluation. |
| EE: Electrical Engineering<br>EES: Environmental and ecological science | Technical—design and implementation |
| INS: Interdisciplinary<br>IS: Internet Science | **Educational Objectives** |
| ME: Mechanical Engineering<br>P: Physic | C.U. Conceptual Understanding<br>D.S.: Design Skills |
| MME: Mechanical and Manufacturing Engineering | P.S.: Professional Skills<br>S.S.: Social Skills |

**Table IV. Simulated Laboratory Article Objectives**

| Art. | Sub. | Meth. | Constraints | | Educational Objectives | | | |
|---|---|---|---|---|---|---|---|---|
| | | | Time & Cost | Acc. & Flex. | C.U. | P.S. | D.S. | S.S. |
| [1] | EE | T | | √ | √ | | √ | |
| [2] | TE | T | | | √ | √ | √ | √ |
| [3] | EE | T | √ | | √ | | √ | |
| [4] | EE | T | √ | √ | √ | | | |
| [5] | ME | T | √ | √ | √ | | | |
| [6] | EE | Q | √ | | √ | √ | | |
| [7] | CE | T | √ | √ | √ | | | |
| [8] | ME | Q | √ | | √ | √ | | |
| [9] | ME | Q | √ | | √ | √ | | |
| [10] | Phi | Q | √ | √ | √ | | | |
| [11] | ME | Q | √ | | √ | √ | | |
| [12] | INS | T | √ | √ | √ | | | |
| [13] | CE | T | √ | √ | √ | | | √ |
| [14] | B | T | √ | √ | √ | | | |
| [15] | Clm | Q | √ | | √ | √ | | |
| [16] | CVE | T | √ | √ | √ | | | |
| [17] | ME | T | √ | | √ | √ | | |
| [18] | PE | T | √ | | √ | | | √ |
| [19] | PE | T | | | √ | √ | | |
| [20] | PE | T | | | √ | √ | | |
| SUM | | | 13 | 11 | 20 | 16 | 9 | 5 |

**Table III. Hands-On Laboratory Article Objectives**

| Art. | Sub. | Meth. | Constraints | | Educational Objectives | | | |
|---|---|---|---|---|---|---|---|---|
| | | | Time & Cost | Acc. & Flex. | C.U. | P.S. | D.S. | S.S. |
| [1] | ME | Q | √ | | √ | √ | | √ |
| [2] | ME | Q | √ | | √ | √ | | |
| [3] | ME | E | | | √ | √ | √ | |
| [4] | ME | Q | √ | | √ | √ | | |
| [5] | EE | Q | | | √ | √ | √ | |
| [6] | MME | Q | √ | √ | √ | | √ | |
| [7] | ME | E | | | √ | √ | √ | |
| [8] | SE | Q | √ | | √ | √ | | |
| [9] | CE | E | | | √ | | √ | |
| [10] | B | Q | | | √ | | | √ |
| [11] | EE | Q | | | √ | √ | √ | |
| [12] | EE | Q | √ | | √ | √ | √ | √ |
| [13] | CE | Q | | | √ | | √ | |
| [14] | P | Q/T | | | √ | √ | | |
| [15] | CE | Q | | | √ | √ | √ | |
| [16] | P | Q | | | √ | | | |
| [17] | AE | Q | | | √ | √ | √ | |
| [18] | EES | Q | | | √ | √ | √ | |
| [19] | CE | Q | | | √ | | | |
| [20] | ME | Q | | | √ | √ | √ | √ |
| SUM | | | 6 | 2 | 20 | 15 | 13 | 8 |

**Table V. Remote Laboratory Article Objectives**

| Art. | Sub. | Meth. | Constraints | | Educational Objectives | | | |
|---|---|---|---|---|---|---|---|---|
| | | | Time & Cost | Acc. & Flex. | C.U. | P.S. | D.S. | S.S. |
| [1] | ME | T | √ | √ | √ | | | |
| [2] | ME | T | √ | √ | √ | | | |
| [3] | EE | T | √ | √ | √ | | | |
| [4] | EE | T | | | √ | √ | | |
| [5] | IS | T | √ | √ | √ | | | |
| [6] | EES | T | √ | √ | √ | | | |
| [7] | EE | T | √ | √ | √ | | | |
| [8] | EE | T | √ | √ | √ | | | |
| [9] | EE | T | √ | √ | √ | | | |
| [10] | EE | T | √ | √ | √ | | | |
| [11] | EE | T | √ | √ | √ | | | |
| [12] | EE | T | √ | √ | √ | | | |
| [13] | SE | Q | √ | √ | √ | | | √ |
| [14] | P&CE | Q | √ | √ | √ | | | |
| [15] | CS | T | √ | √ | √ | | | |
| [16] | EE | T | √ | √ | √ | | | |
| [17] | B | Q | √ | √ | √ | √ | √ | |
| [18] | EE | T | √ | √ | √ | | | |
| [19] | EE | T | √ | √ | √ | | | |
| [20] | EE | T | √ | √ | √ | | | |
| SUM | | | 17 | 19 | 19 | 13 | 1 | 4 |

## II. ARTICLES ON HANDS-ON LABS

## III. ARTICLES ON SIMULATED LABS

## IV. ARTICLES ON REMOTE LABS

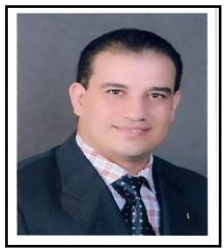

**Yasser . H. Elawady** is a Lecturer in the Department of Computer Engineering, Faculty of Computers and Information Systems, Taif University, Taif, KSA. He received his M.Sc. from the Department of Computer Engineering, Faculty of Engineering, Mansoura university, Mansoura, Egypt, in 2003. His subject of interest includes *Remote Access, FPGA programming, Hardware Design* , *computer architecture and organization and Networking*.

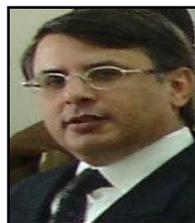

**A. S. Tolba** is a full professor at Mansoura University, Egypt. He is a professor of Computer Science and Engineering. Professor Tolba holds a PhD in Computer Vision from Wuppertal University in Germany.
He is the author of over 70 papers that have been published in refereed international Journals (Computers in Industry, Cybernetics and Systems, Digital Signal Processing, Pattern Analysis and Applications, Biomedical Research, International Journal of Hybrid Intelligent Systems, International Journal of signal Processing, International Journal of Computers and Applications and IJCSNS International Journal of Computer Science and Network Security) and conferences in the areas of Face Recognition, Neural Networks, Glove-Based Gesture Recognition, Speaker Recognition, MRI Compression, Data mining, Automated Visual Inspection of Flat Surfaces, Combined Classifiers, Signature Recognition and eLearning. He has served as the director of the national project "ICT in Higher Education Development in the Egyptian Universities". He is the founder and director of the eLearning Center at Mansoura University, Egypt during the period (2005-2007) and served as the Director of the National eLearning Center (2006-2007).
Professional activities include Deputy Dean of Graduate Studies and Research (2003), Dean of the Faculty of Computer Science and Information Systems in Mansoura University (2004-2007) and he is currently, the Dean of the Faculty of Computer Studies at the Arab Open University, Kuwait. He served as an IT and Educational Technology consultant for the minister of general education in Egypt in 2006. He served also as a reviewer in many international Journals: IEEE Transaction on Pattern Analysis and Machine Intelligence, Image Processing and Vision Computing. He has published books in Computer Vision and Robotics, Edited Book, 1990 (Kluwer Academic Publishers), Lasers in Computing and Health (ALESCO, Morocco, 1997), ICTs and Higher Education in Africa,2007, Published under a Creative Commons License, 2007, and E-learning in General Education, 2008 (The Arab network for Open and Distance Education, National Library, Jordan).